\documentclass[runningheads]{llncs}
\usepackage[T1]{fontenc}
\usepackage{graphicx}
\usepackage{cite}
\usepackage{amsmath,amssymb,amsfonts}
\usepackage{algorithmic}
\usepackage{graphicx}
\usepackage{textcomp}
\usepackage{multirow}
\usepackage{xcolor}
\usepackage{tikz}
\usepackage{tabularx}
\usepackage{hyperref}
\begin{document}
\title{Unsupervised User-Based Insider Threat Detection Using Bayesian Gaussian Mixture Models}
%
%
\author{Simon Bertrand\orcidID{0000-0002-8518-0770} \and
Josée Desharnais\orcidID{0000-0003-2410-3314} \and
Nadia Tawbi\orcidID{0000-0002-1030-0918}}
\authorrunning{S. Bertrand et al.}
%
\institute{Département d'informatique et de génie logiciel\\
Laval University, Québec Qc, Canada\\
\email{siber93@ulaval.ca, \{josee.desharnais,nadia.tawbi\}@ift.ulaval.ca}}
\maketitle              
\begin{abstract}
Insider threats are a growing concern for organizations due to the amount of damage that their members can inflict by combining their privileged access and domain knowledge. Nonetheless, the detection of such threats is challenging, precisely because of the ability of the authorized personnel to easily conduct malicious actions and because of the immense size and diversity of audit data produced by organizations in which the few malicious footprints are hidden.  
In this paper, we propose an unsupervised insider threat detection system based on audit data using Bayesian Gaussian Mixture Models. The proposed approach leverages a user-based model to optimize specific behaviors modelization and an automatic feature extraction system based on Word2Vec for ease of use in a real-life scenario. The solution distinguishes itself by not requiring data balancing nor to be trained only on normal instances, and by its little domain knowledge required to implement. Still, results indicate that the proposed method competes with state-of-the-art approaches, presenting a good recall of 88\%, accuracy and true negative rate of 93\%, and a false positive rate of 6.9\%. 
For our experiments, we used the benchmark dataset CERT version 4.2.

\keywords{Insider Threat \and Bayesian Gaussian Mixture Model \and Gaussian Mixture Model \and Unsupervised \and Word2Vec}
\end{abstract}
\section{Introduction}
Insider threats occur when a privileged member of an organization wrongfully uses their access in a way that causes harm to their organization. Those damaging actions can be intentional, as in the case of theft or sabotage, however, unintentional dangerous actions are also to be considered, which adds to the complexity of the insider threat. The unintentionally dangerous insider is mostly acting by negligence or misinformation. An example is an employee who copies corporate sensitive data locally by convenience. Such actions, even if seen as negligible, can compromise the confidentiality of the data by creating a second access point to the information that can be exploited by a hacker. The insider threat is then a broad type of cyber menace which makes its detection particularly difficult.

For organizations, the confidentiality, the integrity, and the availability of their information are an increasing concern. Yet many underestimate the magnitude of the insider threat against the maintenance of those ideals. Indeed, even though insider threats are only a fraction of all existing cyber threats, this type of menace presents a real and unique danger for organizations. Firstly, the insider threat can be more damaging to an organization than a traditional cyberattack. This is mainly explicable by the privileged accesses and great domain knowledge that the insider possesses over an outsider. The insider has then a better opportunity to use their access and domain knowledge to carry out efficiently and quietly the attack. Moreover, over the last years, some reports suggest that most institutions suffer from that kind of cyber threat yearly. For instance, according to the "Insider Threat Report" of 2019, about 60\% of organizations were victims of at least one insider attack in 2019 \cite{more}.

Insider threat detection is then a relevant problem that attracted many researchers to deploy their efforts in the last decades. One common strategy to detect malicious insiders is by modeling the behaviors of the users and identifying any significant divergence as a potential threat. In that matter, audit data, describing the activity of every member of an organization in the network, are regularly chosen to learn user behaviors using statistical or machine learning models.

However, detecting insider threats based on audit data presents many challenges, one of which is to efficiently consider sequence information to learn behaviors. Indeed, like most cyber threats, an insider attack is rarely defined by a single malicious event, but mostly as a series of events. Additionally, not only can the malicious series of events be scattered over a period of time, but they are also often sequence dependent, meaning that the order in which the events occur is important to identify if the sequence is malicious or not. For instance, considering a simple data exfiltration threat, the event of reading sensitive data before writing an email is more suspicious than the other way around. Few existing works, based on machine learning, focus on using the event sequence information in the behavior learning process on long time windows.

In addition, another challenge is to create a solution that is flexible and adapted for real-life organizations. This challenge is mostly due to the singularity and complexity of all organizations' technology architecture, which leads to a lack of public datasets that represent the reality of all organizations. While using public datasets is convenient for comparison purposes, one needs to be careful when processing the data from such datasets. One risk is to overfit a solution to a specific dataset which can lead to the solution performing poorly in other settings. For instance, to use the label information in the dataset can be convenient to balance the data classes or extract positive instances for One-Class training. However, relying on such information is an issue when implementing the solution in a real-world case where organizations rarely possess historically labeled audit data. Furthermore, in some cases, organizations can even difficultly guarantee that historical audit data are threat free. Those limitations make supervised solutions unsuitable for organizations and highlight a need for unsupervised alternatives.

\section{Contributions}
Considering those challenges, in this paper, we propose an insider threat detection system that uses unsupervised machine learning, trained on processed audit data, to detect malicious conduct. More precisely, the proposed technique consists of training a Bayesian Gaussian Mixture Model (BGMM) for every user, utilizing their historical audit data to learn normal behavior clusters/components. By using a user-based framework, not only do we differentiate between users, but we can take advantage of the  BGMM's ability to determine an effective number of components for every single user, which allows for an adapted model that fits the different types of behaviors of a specific user. In addition, to address the challenge of sequences dependencies of insider threats, and to facilitate feature extraction, a user-based Word2Vec model is trained to capture contextual information about the activities in the host logs to generate a daily activity summary vector. The proposed solution is then a combination of deep learning models used for data pre-processing and statistical models to learn user behaviors. 

A user-based model is uncommon among existing insider threat detection techniques which motivates our efforts to research for improvements in that field. In that matter, we explore the effects of a custom number of clusters/components for every user on insider threat detection performances, which is the main thesis of this paper. There is, to our knowledge, no existing work that combined an automatic number of clusters/components in a user-based setting.

Furthermore, to deal with the flexibility challenge, the proposed method is developed with the restriction of requiring as little domain knowledge and data pre-processing as possible, thus increasing its flexibility. In that matter, the proposed method doesn't require data balancing nor to be trained only on normal instances which we believe reflects as fairly as possible the reality of real organizations. In addition to those characteristics, we believe that the simplicity in the selection of features also contributes to making the solution more suitable for a real-life scenario.

Despite those restrictions, the proposed approach outperforms state-of-the-art methods in the accuracy, false positive rate, and true negative rate metrics, and still offers a competitive recall rate.

To sum up, the proposed solution stands out because of the following characteristics:
\begin{itemize}
  \setlength\itemsep{0.75em}
  \item Completly unsupervised.
  \item User-based framework.
  \item Automatic feature extraction with Word2Vec models.
  \item Automatic selection of an efficient number of components per user with BGMM models.
  \item Low on domain knowledge.
  \item No data balancing.
  \item Beats relative state-of-the-art techniques.
\end{itemize}
The rest of this paper is organized as follows. Section 3 presents previous research concerning insider threat detection. Section 4  introduces the proposed method, while section 5 presents experimental results and a comparison of the proposed method with similar state-of-the-art techniques. Finally,  section 6 presents a conclusion containing potential improvements for future work.

\section{Related Work}
Previous work in the field of unsupervised insider threat detection based on audit data can principally be grouped into two categories: signature-based and machine learning based techniques.

Signature-based threat detection techniques mainly consist of the creation of a dictionary of allowed and/or disallowed activity patterns. The dictionary is then prompted to check if any sequence of activity matches those in the lexicon to determine if the behavior is normal or abnormal. Signature-based techniques offer the advantage of a low false positive rate, but requires frequent updates to detect new anomalies \cite{Khraisat2019SurveyOI} which requires manually introducing the new threats in the dictionary. 

Machine learning techniques present an appealing alternative to signature-based methods because of their ability to learn automatically normal and abnormal behaviors. This feature generally increases their flexibility and reduces the required domain knowledge \cite{10.1007/978-3-030-36938-5_2}. In the last decades, many machine learning based insider threat detection systems were constructed on statistical or clustering techniques. For instance, Eldardiry et al. \cite{6565228} propose a user-based model system that uses the K-Means algorithm to model daily user behavior. Happa \& Tabash \cite{Happa} presente a similar framework but Gaussian Mixture models (GMMs) are used to detect anomalous instances by selecting the data points that are the less likely to have been generated by the learned Gaussian distributions. Kim et al. \cite{app9194018} propose a study on the existing clustering and statistical techniques presenting the performances of K-Means, Parzen Window Density Estimation, Principal Component Analysis, and Gaussian algorithms. In general, statistical and clustering techniques offer a simple way to detect the malicious insider, but suffer from a high false positive rate \cite{6133296}.

A leap in the field of unsupervised insider threat detection has been the use of Autoencoders. Zhang et al. \cite{DA} propose the use of compressed daily feature representations obtained with a Denoising Autoencoder as input to a GMM, to learn normal behaviors and detect divergent instances. The suggested solution stands out from other techniques partly because of the use of a Word2Vec model for automatic feature extraction, which inspired part of this work.

Finally, with the increasing popularity of deep learning in recent years, many researchers integrated the use of deep models to the detection of the insider threat issue. Many RNN-based methods have been proposed to detect insider threats because of their ability to model time series data. However, the conventional RNN tends to perform poorly in long time series \cite{8478898}, which can be problematic in audit data based anomaly detection tasks due to the long sequences of events a user usually performs daily. This limitation is why most current efforts in the deep learning field use LSTM neural networks which are known to capture long-time dependencies \cite{8411899}. As examples,  Nasir et al. \cite{nasir} and Sharma et al.\cite{10.1145/3406601.3406610} propose the use of LSTM-AutoEncoders to encode and decode session activity summary vectors and label the sessions with the highest reconstruction error score as being of malicious nature. Results demonstrate that deep learning methods provide overall better results compared with shallow machine learning models, specifically a lower false positive rate.

\section{Methodology}
The proposed solution consists in learning the daily behaviors of every user in the organization and detecting any day diverging significantly from the typical user behavior as being anomalous, thus probably containing at least 1 malicious event. To do so, a feature extraction process that necessitates as little domain knowledge as possible, and user-specific behavior learning models are used.

\begin{figure}
\centering
\centerline{\includegraphics[width=0.9\textwidth]{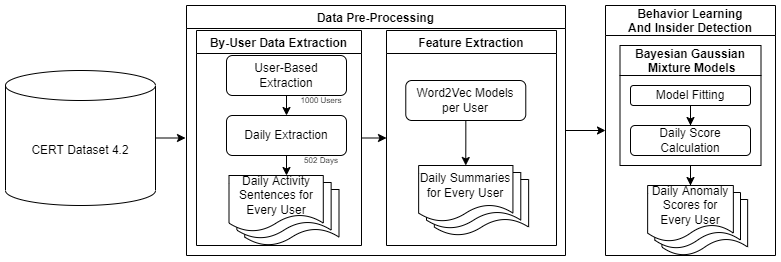}}
\caption{Proposed Framework}
\label{fig1}
\end{figure}

The proposed framework is presented in Fig.~\ref{fig1}. The framework is separated into a data pre-processing phase and a behavior learning/insider detection phase.

\subsection{Data Collection}
The dataset used in this study is the commonly used CERT insider threat dataset version 4.2 \cite{CERT} created by Carnegie Mellon's Software Engineering Institute \cite{6565236}. This synthetic dataset is composed of audit data generated by simulating a 1000 employees organization within 502 days. The audit data is separated into 5 domain types: logon, emails, files, HTTP, and devices. Each domain contains one or two specific activity types. All the activities in the Dataset count for 32,770,227 events in total. Table~\ref{tab1} 
\begin{table}
\renewcommand{\arraystretch}{1.5}
\caption{Dataset Activities Description}
\label{tab1}
\centering
\begin{tabular}{c|c|l}
\hline
\textbf{Domain} & \textbf{Activity} & \textbf{Description} \\
\hline
\text{Logon} & \text{Logon} &\text{Connection using the userid }\\
\hline
\text{Logon} &\text{Logoff} & \text{Disconnection of the userid}\\
\hline
\text{HTTP} &\text{HTTP} & \text{Website access}\\
\hline
\text{Email} &\text{Email} &\text{Creation of an email}\\
\hline
\text{File} &\text{File} &\text{File-level access} \\
\hline
\text{Device} &\text{Connect} &\text{Connection of a USB device} \\
\hline
\text{Device} &\text{Disconnect} &\text{Disconnection of a USB device} \\
\hline
\end{tabular}
\end{table}
presents the different domains and activity types. Each activity domain audit data is stored in a CSV file. The dataset is mainly composed of normal events, but malicious activities perpetrated by 70 users were injected and account for only 0.03\% of the dataset instances. In that regard,  one challenge regarding the use of this dataset is the fact that the normal and malicious instances are imbalanced. Even though balancing efforts have been proven, in other works, to be beneficial for the detection of malicious insiders \cite{8962316}, in this work no data balancing is performed. This decision is motivated by our goal to create a solution that is more appropriate for a real-world scenario, where it is difficult to have any knowledge about past anomalous actions.

\subsection{Data Pre-processing}
\subsubsection{User Behavior Data Extraction}
Using the previous dataset, the activity in the 5 domains is firstly aggregated for every single user. This aggregation allows to easily get the activity of a user across all domains. Furthermore, having a user-based model configuration, this data grouping is essential.

 The user aggregated data is then grouped by day, so every instance represents the daily activities of an individual for a total of 330 452 instances, with only 966 having any malicious activities occurring during the day. In this work, we only consider the activity type itself and the order in which it occurs to create the behavioral model, so every other information is ignored. Motivated by our objective to create an easy-to-implement solution, the choice to keep only the activity type is driven by its standardness and little domain knowledge required to identify. The results of this step are strings object composed of event types carried out by a user daily.

\subsubsection{Feature Extraction}
One of the difficulties in generating feature vectors representing a condensed summary of a time window is capturing the temporality and dependencies of events. In that matter, we explore the use of Word2Vec \cite{https://doi.org/10.48550/arxiv.1301.3781} models. Using the Skip-gram model, Word2Vec can capture syntactic and semantic word relationships automatically \cite{https://doi.org/10.48550/arxiv.1310.4546}.   Therefore, in this work, this model's purpose is to capture the user's daily specific behavior patterns like the typical order of the activities. Specifically, a Word2Vec model is trained for every user using their daily activity strings to learn context-rich word embeddings for every event. This step ensures that every user has custom word embeddings depending on their behaviors. In addition to its ability to generate context-rich word embeddings, using Word2Vec models contributes to our objective to reduce the dependency of the domain knowledge \cite{DA}, by automating the extraction of complex features from simple data, being the activity type.

 Finally, using the trained Word2Vec models, every user's daily activities are summarized into a vector for every day they were active in the organization. To do so, the Word2Vec model receives one by one the daily activities of a user and transforms every single activity into an embedding that is then summed with the other activities embedding that occurred during that same day. The resulting vector is then a daily summary that has information about the volume and type of events carried out and contains contextual details about the order of execution of the user's daily activities. It is important to note that the embeddings are generated from past data, meaning that if used in a real-time scenario, the new sequences will not have an impact on the embeddings generated.

\subsection{Behavior Learning}
In the pre-processing phase, daily activity summary vectors containing contextual information are extracted for every user. To learn the user's normal behaviors using those vectors, we explore GMMs for their high recall rate \cite{DA}. Furthermore, assuming that a user can have more than one normal behavior, we rely on the GMM's ability to learn from multimodal data distributions to fit clusters/components representing behaviors. Moreover, because of the low volume of malicious records in the dataset, our intuition is that every component mostly describes a normal behavior for the user, and thus high-density regions in a component describe normal behaviors, and low-density ones abnormal behaviors. In other words, we deal with the detection of abnormal behaviors as an outlier detection task.

However, GMMs present the challenge of the selection of the right number of components. For that reason, in this work, we decide to explore the effects of an automatic solution for the selection of the number of components in a user-based framework. Leveraging the fact that the solution is user-based, the intuition is that we could optimize the performance by selecting the right number of components for every user versus a global number of components for all users.

This intuition comes from the fact that every user has different and unique normal behaviors. Consequently, because in a GMM every component is likely to represent a behavior, choosing the right number of Gaussian distributions for every user could result in better modeling and anomaly detection power. A simple example could be a user that exhibits two normal behaviors, the first one being their email intensive days, maybe occurring early on in the week and the second one being characterized by higher file and website activities. For this particular user, a 2 component model would probably suffice. A higher number of components, in this case, could mean overfitting the data. In an anomaly detection task using partitioning models, overfitting a model can lead to the creation of a separate cluster, grouping the anomalies and making them appear normal by most distance or density metrics. However, for another user having a specific behavior depending on the day of the working week, a 5 component model could be more appropriate. In that last example, choosing a lower number of components could mean underfitting the data, and thus overgeneralizing the behaviors.

Many options are available to optimize the number of components of GMMs, like the Bayesian Information Criterion (BIC) score and the Akaike Information Criterion (AIC) score, but in this work, we select a variation of the GMM: the \emph{Bayesian Gaussian Mixture Model} which is fundamentally a GMM but using the \emph{Dirichlet} process to infer a weight of importance for every component. This technique only requires that a likely superior number of components be provided as an input. Irrelevant components will be assigned a weight of 0.

To understand the \emph{Bayesian Gaussian Mixture Model}, it is important to first present the \emph{Gaussian Mixture Model}.\\
\subsubsection{Gaussian Mixture Model}
The GMM is a probabilistic model often used in anomaly detection tasks due to its high recall rate \cite{DA}. The GMM is composed of a set of parameters which are described in Table~\ref{tab2}.

\begin{table}
\renewcommand{\arraystretch}{1.5}
\caption{Gaussian Mixture Model's Parameters}
\begin{center}
\begin{tabularx}{\linewidth}{c|l}
\hline
\textbf{Parameter} & \textbf{Description}\\
\hline
\text{K} &\text{Number of Gaussian distributions (number of components)}\\
\hline
\text{N} &\text{Number of instances}\\
\hline
\text{$\phi_{i=1...K}$} &\text{Prior probability of component i}\\ 
\hline
\text{$\mu_{i=1...K}$} &\text{Mean of component i}\\ 
\hline
\text{$\Sigma_{i=1...K}$} &\text{Covariance of component i}\\ 
\hline
\text{$\phi$} &\text{K-dimensional Prior probability vector}\\ 
\hline
\text{$\mu$} &\text{K-dimensional mean vector}\\ 
\hline
\text{$\Sigma$} &\text{K x K covariance matrix}\\ 
\hline
\text{$\mathcal{N}$} &\text{Gaussian distribution}\\ 
\hline
\end{tabularx}
\label{tab2}
\end{center}
\end{table}
In this model, we assume that the data obey a mixture of several Gaussian distributions defined as:\\
\begin{equation}
\mathcal{N}(x|\mu_{i},\Sigma_{i}) = \frac{1}{\sqrt{{(2\pi)}^{K}|\Sigma_{i}|}}\exp{-\frac{1}{2}(x - \mu_{i})^{T}\Sigma^{-1}_{i}(x - \mu_{i})}.
\end{equation}

For instance, when analyzing a user's daily activity summaries, a Gaussian distribution could explain email intensive days and another Gaussian distribution could explain file intensive days.

The GMM, knowing the number of components, learns to fit the instances in a way that maximizes the log-likelihood of the dataset, defined as:\\
\begin{equation}
\log{p(x| \mu, \phi, \Sigma)} = \sum_{n=1}^N\log{\sum_{i=1}^{K}\phi_{i}\mathcal{N}(x|\mu_{i},\Sigma_{i})}.
\end{equation}

To do so, the Expectation Maximization algorithm \cite{10.2307/2984875} learns the $\mu$, $\phi$ and $\Sigma$ parameters that maximize Equation 2 for the dataset, which is the sum of every instance's log probability.

The probability of a single instance having been generated by the mixture model is calculated as follows: \\

\begin{equation}
p(x)=\sum_{i=1}^{K}\phi_{i}\mathcal{N}(x|\mu_{i},\Sigma_{i}),
\end{equation}

which is simply the sum of the probability of the instance to be part of each Gaussian distribution multiplied by the prior probability or the weight of the component. \\

So the log-likelihood is calculated as follows:
\begin{equation}
\log{p(x)}=\log{\sum_{i=1}^{K}\phi_{i}\mathcal{N}(x|\mu_{i},\Sigma_{i})}.
\end{equation}

Even though GMMs are often used in anomaly detection tasks, the choice of a good number of components can be a difficult chore. Our intuition is that the optimal number of components will vary from user to user, because user's behaviors are unique. Therefore, a potential gain can be achieved by having a per-user model for which a custom number of components is selected.\\
\subsubsection{Bayesian Gaussian Mixture Model}
In this work, a variation of GMMs, the BGMM is explored. This variation can infer an efficient number of components from the data. The BGMM is very similar to a GMM except for the fact that it is a non-parametric model, meaning it uses variational inference to estimate the model parameters. This variation requires the use of prior distributions over the parameters of the GMM. Then the parameters optimization process follows the Expectation Maximization algorithm, but computes the entire posterior distribution over the parameters for regularization.

Precisely, the proposed BGMM uses the same parameters as described in table II, but with a Dirichlet process as a distribution to infer the number of components $\phi$.

To understand the Dirichlet process, the stick-breaking analogy can be used. Imagining the prior weights of our unknown number of components as a long stick with a length of 1, the Dirichlet process consists of breaking the stick continually and associating the points that fall into a group of a mixture. For each division of the stick, new Gaussian distributions or components are created, having for weight the length of the piece of the stick. The stick can be broken down infinitely, thus generating an infinite number of components. From that division, a good number of components can be assessed by removing components with negligible weights and thus avoiding overfitting. Because of the infinite nature of the Dirichlet process, and the inability of computers to deal with infinity, the model used in this solution still requires to give an upper-bound number of components.

\subsection{Insider Threat Detection}
After the learning process, a score for every user's day is calculated using the trained BGMM. The score represents the log-likelihood, presented in Equation 4, of the instance compared to its model. A low score means that the day is at the furthest end of our learned Gaussian Distributions or/and is part of a lower weight component, which we interpret as abnormal behavior. Every day's score is divided by the mean of the user's scores to get a ratio of that day's score against what is normal for the user. The reason for that ratio is because a normal score varies from one user to another and so using the ratio will tell how far from its mean the day is, which can be compared across users. Finally, with the daily score ratios of every user, every day that has a greater score to a threshold is identified as malicious.

\section{Implementation and Results}
In this section, we evaluate the performance of our solution. We first compare it with similar state-of-the-art techniques. Then, we verify our assumption that a custom number of components for every user is beneficial. Finally, we check if the performance varies between executions.

The implementation is done using an Ubuntu 20.04.4 LTS operating system with an Intel(R) Core (TM) i7-8700 CPU @ 3.20GHZ X 12 cores and 16 GB of RAM. The language used is Python and is executed on a jupyter notebook. Scikit-learn's \cite{sklearn_api} \emph{Bayesian Gaussian Mixture Model} and Gensim's \emph{Word2Vec} libraries are used. 100\% of the data is used for training and detection in an unsupervised way. The proposed method doesn't require data balancing or to be trained only on normal instances.

To evaluate our solution we use the false positive rate ($\mathrm{FPR}$), recall, true negative rate ($\mathrm{TNR}$), and accuracy metrics, being common metrics to evaluate anomaly detection models. The metrics formulae are as follows, where a negative instance refers to a normal day and a positive to a day containing at least one malicious event:
\begin{align}
\mathit{FPR}&=\mathit{FP}/(\mathit{TN}+\mathit{FP})
\\[1mm]
\mathit{Recall}&=\mathit{TP}/(\mathit{TP}+\mathit{FN})
\\[1mm]
\mathit{TNR}&=\mathit{TN}/(\mathit{TN}+\mathit{FP})
\\[1mm]
\mathit{Accuracy}&=(\mathit{TP} + \mathit{TN})/(\mathit{TP} +\mathit{TN} + \mathit{FP} + \mathit{FN})
\end{align}

Where $\mathrm{TN}$ is true negative, $\mathrm{FP}$ is false positive, $\mathrm{TP}$ is true positive and $\mathrm{FN}$ is false negative.

We use the preceding metrics to compare how the proposed method performs comparatively with other relevant techniques using the same dataset. Table~\ref{tab3} 

\begin{table}
\caption{Results Comparison}
\begin{center}
\renewcommand{\arraystretch}{2}
\begin{tabular}{|c|c|c|c|c|c|}
\hline
\textbf{Method} & \textbf{Reference} & \textbf{Recall}& \textbf{FPR}& \textbf{Accuracy} & \textbf{TNR}\\
\hline
\text{Isolation Forest} & \text{\cite{Gamachchi2017GraphBF}} & \text{NA}& \text{NA}& \text{79\%} & \text{NA}\\
\hline
 \text{LSTM-AutoEncoder} & \text{\cite{10.1145/3406601.3406610}} & \textbf{91.03\%}& \text{9.84\%}& \text{90.17\%} & \text{90.15\%}\\
\hline
 \text{DBN-OCSVM} & \text{\cite{8117081}} &\text{81.04\%}& \text{12.18\%}& \text{87.79\%} & \text{NA}\\
\hline
\text{DA With Clustering} &  \text{\cite{DA}} &\text{88.9\%}& \text{20\%}& \text{75\%} & \text{NA}\\
\hline
\text{Proposed}  & \text{NA} &\text{88.38}& \textbf{6.9\%}& \textbf{93.08\%} & \textbf{93.10\%}\\
\hline
\end{tabular}
\label{tab3}
\end{center}
\end{table}
presents a comparative table where the selected metrics for each method are displayed if it is available in the corresponding paper. For the proposed approach, the results were obtained by calculating the mean of 100 random executions, for validity.

According to the results presented in Table 3, we can see that the proposed approach outperforms other comparative techniques on the accuracy, false positive rate, and true negative rate metrics and still presents a competitive recall. Furthermore, it is important to mention that the presented results were achieved even if the proposed method does not rely upon more domain knowledge intensive attributes, data balancing, or one-class training, as opposed to other techniques.

Table~\ref{tab4}
\begin{table}
\caption{Confusion Matrix}
\label{tab4}
\begin{center}
\renewcommand{\arraystretch}{2}
\begin{tabular}{cc|cc}
\multicolumn{1}{c}{} &\multicolumn{1}{c}{} &\multicolumn{2}{c}{Predicted Class} \\ 
\multicolumn{1}{c}{} & 
\multicolumn{1}{c|}{} & 
\multicolumn{1}{c}{Insider} & 
\multicolumn{1}{c}{Normal} \\ \hline
\multirow[c]{2}{*}{\rotatebox[origin=tr]{90}{True Class}}
& \phantom{iiii}  Insider  & 863 & 106   \\[1.5ex]
& \phantom{iiii}  Normal  & 22\,960   & 306\,526 \\ \hline
\end{tabular}
\end{center}
\end{table}
presents the confusion matrix of one execution of the solution. A strong diagonal can be observed, with 860 days containing malicious activities accurately detected out of 966 in total (for a recall of 89\% for this particular execution). Most normal days are correctly identified, with 306 526 days accurately predicted out of 329 486. Even though 22 960 days were wrongfully labeled as anomalous, which is quite high, it is a common problem with unsupervised anomaly detection methods. Nonetheless, a  false positive rate of around 7\%  still outperforms similar state-of-the-art techniques. Furthermore, in this particular field, it is preferable to have a higher false positive rate than to miss out on falsely labeled benign attacks.

We present the receiver operating characteristic or ROC curve to evaluate the performance of our binary classifier. This curve shows the recalls and false positive rates obtained at different threshold configurations. This representation can be a good way for cyber analysts with different time budgets to find a good compromise by choosing a threshold that doesn't lead to an unrealistic amount of false positives and offers an expected recall rate with which they are comfortable. Because it can be hard to draw valid conclusions from a ROC curve,  the area under the ROC curve (\emph{AUC}) is a metric that can be used to globally evaluate and compare a model's performance.

\begin{figure}
\centering
\centerline{\includegraphics[width=110mm]{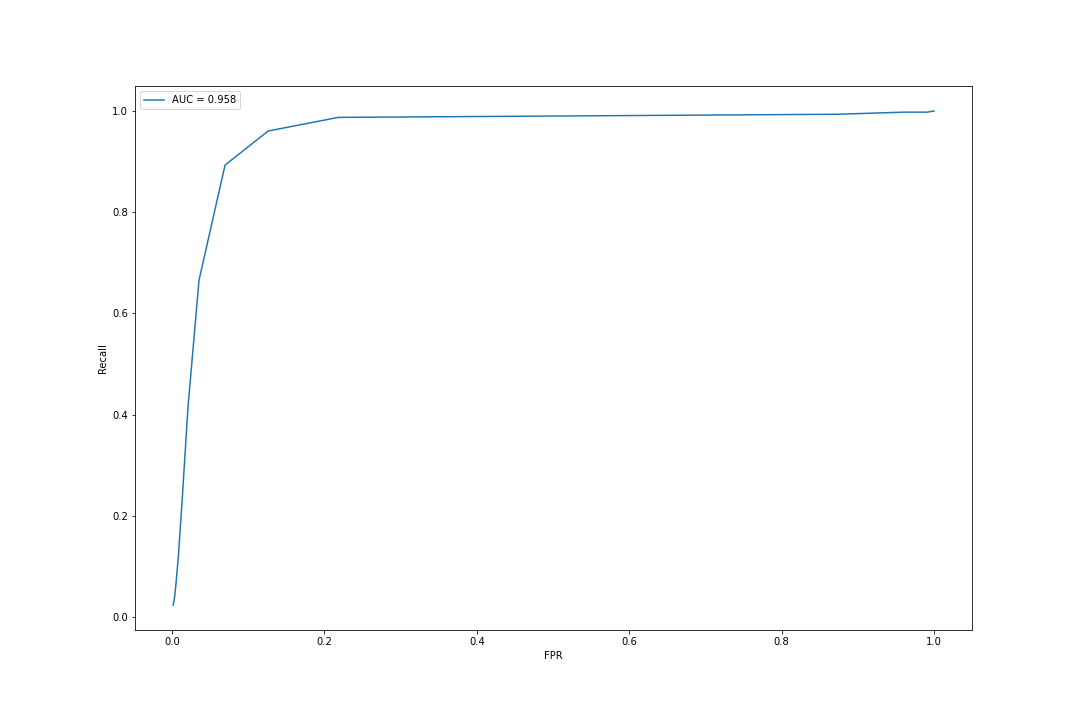}}
\caption{ROC Curve}
\label{fig2}
\end{figure}

Fig.~\ref{fig2}
 presents the ROC curve of an instance of the proposed method. The curve shows that the proposed solution offers many interesting opportunities for insider threat detection depending on an organization's available resources. For instance, an organization with a low budget for the analysis of instances could use a low threshold to minimize the false positive rate, thus reducing the total amount of instances to analyze and still capturing a good portion of all anomalies. The proposed solution presents an \emph{AUC} score of  0.958. To compare with another state-of-the-art method using the same dataset, an \emph{AUC} score of 0.949 is achieved by Sharma et al. \cite{10.1145/3406601.3406610}.

\begin{figure}
\centering
\centerline{\includegraphics[width=110mm]{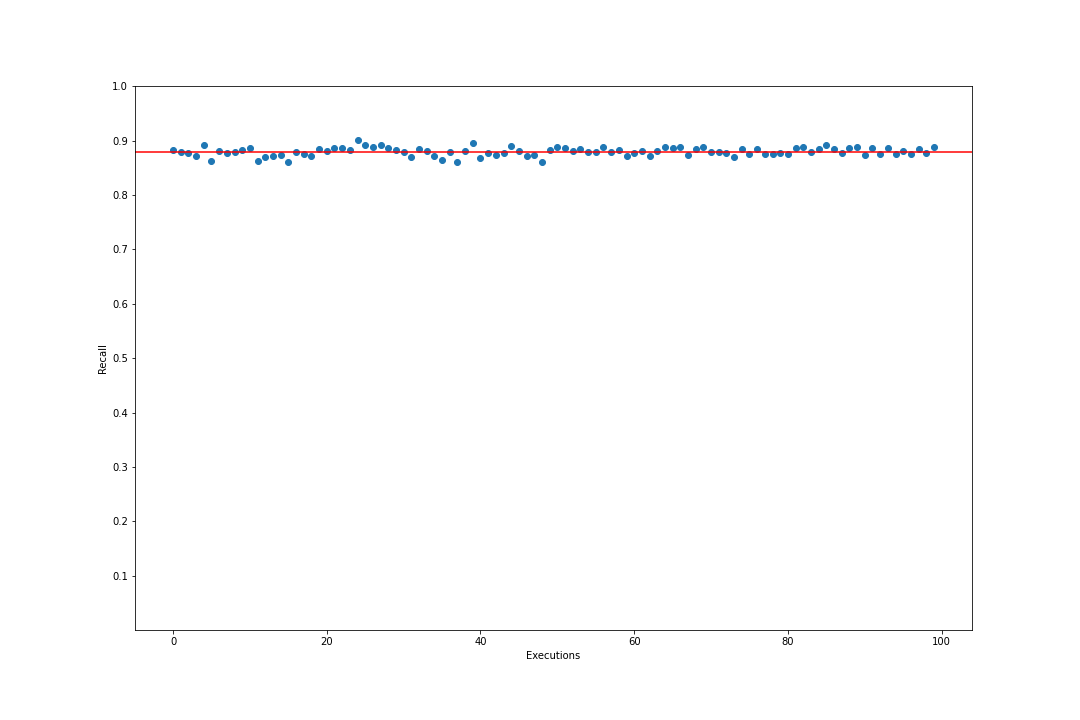}}
\caption{Recall of 100 random executions. The mean is identified in red.}
\label{fig3}
\end{figure}

Because a machine learning model's performance can vary due to its random elements, we wanted to see if the proposed solution was sensitive to randomness.  Fig.~\ref{fig3}
presents the recall of the 100 random executions of the solution. This figure shows that the performance doesn't vary greatly and so is not substantially affected by randomness. This means that for any execution, one can assume to have performances close to the presented mean.

In this study, we also wanted to verify the intuition that performance could be optimized by a per-user \emph{custom} number of components. To do so, we compare our results with three configurations of the traditional GMM in Table~\ref{tab5}..
\begin{table}
\caption{Fixed versus custom number of components}
\begin{center}
\renewcommand{\arraystretch}{2}
\begin{tabular}{|c|c|c|c|c|}
\hline
\textbf{METHOD} & \textbf{Recall}& \textbf{FPR}& \textbf{Accuracy} & \textbf{TNR}\\
\hline
\text{GMM-1} & \text{78.05\%}& \text{9.02\%}& \text{90.94\%} & \text{90.98\%}\\
\hline
\text{GMM-3} & \text{78.88\%}& \text{8\%}& \text{91.95\%} & \text{92\%}\\
\hline
\text{GMM-5} & \text{78.05\%}& \text{7.28\%}& \text{92.68\%} & \text{92.71\%}\\
\hline
\text{BGMM}  & \textbf{88.38}& \textbf{6.9\%}& \textbf{93.08\%} & \textbf{93.10\%}\\
\hline
\end{tabular}
\label{tab5}
\end{center}
\end{table}
All the pre-processing is the same as the proposed solution, only the model itself is changed for a GMM. Even the score function is identical because, as seen earlier, the GMM and the BGMM are extremely similar and only differ in their parameters learning steps. The 3 configurations are a GMM for every user with 1, 3, and 5 components.

Results suggest that a custom number of components for every user is beneficial, beating every other fixed number of component configurations in every metric.

\section{Conclusion}
In this paper, an unsupervised insider threat detection model is proposed and tested on the benchmark CERT dataset. The dataset is grouped by user and by day with only the ordered activity type kept. Word2Vec model is used to generate user-specific activity embeddings, capturing activity sequence information. BGMMs are finally used with the daily summary vectors to train and detect malicious behaviors. The proposed method is competitive with state-of-the-art techniques without requiring data balancing or being trained only on normal data, all of which with minimal domain knowledge required, which is in line with or objective of creating a flexible solution. Even though the proposed method performs well using only the activity type and its order, we believe that further improvements could be made by integrating other relevant information like the time of the activity, the weekday, and the content of the activities. Furthermore, in future work, we would like to improve the daily embedding generation process in a way that is more suitable for real-time execution. Finally, we would also like to explore the GMM's performance in a real-time setting, and precisely study the inference of the number of components in an evolving environment.

%
%
%
%
\bibliographystyle{splncs04}
\bibliography{Unsupervised_User-Based_Insider_Threat.bib}

\end{document}